\documentclass{amsproc}

\usepackage{hyperref}

\theoremstyle{definition}

\theoremstyle{remark}

\numberwithin{equation}{section}

\begin{document}

\title[Survey]{Survey of mathematical foundations of QFT and perturbative string theory}

\author{Hisham Sati}
\address{}
\curraddr{
Department of Mathematics,
University of Pittsburgh,
139 University Place,
Pittsburgh, PA 15260
}
\email{hsati@pitt.edu}
\thanks{}

\author{Urs Schreiber}
\address{}
\curraddr{
Department of Mathematics,
Utrecht University,
Budapestlaan 6, 3584 CD Utrecht,
The Netherlands
}
\email{urs.schreiber@gmail.com}
\thanks{}

\subjclass[2010]{Primary 
81T40;  secondary 81T45, 81T30, 81T60, 81T05, , 57R56,
 70S05,  18D05,   55U40,  18D50, 55N34, 19L50, 53C08.\\
{\it Keywords and phrases.}
Topological field theory, conformal field theory,
supersymmetric field theory, axiomatic quantum field theory,
perturbative string theory, 
 conformal nets, monoidal categories, higher categories, generalized
cohomology, differential cohomology, quantization, operads.
}

\date{}

\begin{abstract}
  Recent years have seen noteworthy progress in the mathematical formulation of quantum field theory and perturbative string theory. We give a brief survey of these developments. It serves as an introduction to the more detailed collection \cite{SatiSchreiber}.
\end{abstract}

\maketitle

The history of theoretical fundamental physics is the story of a search
for the suitable mathematical notions and structural concepts that naturally model 
the physical phenomena in question.
It may be worthwhile to recall a few examples:
\begin{enumerate}
\item  the identification of symplectic geometry as the underlying structure of
classical Hamiltonian mechanics;
\item the identification of (semi-)Riemannian differential geometry as the
underlying structure of gravity;
\item the identification of group and representation theory as the underlying 
structure of the zoo of fundamental  particles; 
\item the identification of Chern-Weil theory and differential cohomology
as the underlying structure of gauge theories.
\end{enumerate}
All these examples exhibit the identification of the precise mathematical language
that naturally captures 
the physics under investigation. 
While each of these languages upon its introduction into theoretical 
physics originally met with some skepticism or even hostility, we do know in
retrospect that the modern 
insights and results in the respective areas of theoretical physics would have been
literally unthinkable 
without usage of these languages. A famous historical example
is the Wigner-Weyl approach and its hostile dismissal from
mainstream  physicists of the time (``\emph{Gruppenpest}'');  we now know that group theory and representation theory
have become indispensible 
tools for every theoretical and mathematical physicist.

Much time has passed since the last major such formalization success in theoretical
physics. The rise of 
quantum field theory (QFT) in the middle of the last century and its stunning
successes, despite its 
notorious lack of formal structural underpinnings, made theoretical physicists
confident enough to 
attempt an attack on the next open structural question -- that of the quantum
theory of gauge 
forces including gravity -- without much more of a structural guidance than the
folklore of 
the path integral, however useful that had otherwise proven to be. 

While everyone involved readily admitted that nobody knew the full answer to 
\begin{quote}
  \it What is string theory?
\end{quote} 
%
perhaps it was gradually 
forgotten that nobody even knew the full answer to 
\begin{quote}
  \it What is quantum field theory?
\end{quote}
While a huge discussion ensued on the ``\emph{landscape}'' moduli space of
backgrounds for string theory, 
it was perhaps forgotten that nobody even had anything close to a full answer to 
\begin{quote}
  \it What is a string theory background?
\end{quote}
or even to what should be a simpler question:
\begin{quote}
  \it What is a classical string theory background?
\end{quote}
which in turn is essentially the question:
\begin{quote}
  \it What is a full 2-dimensional $\sigma$-model conformal field theory?
\end{quote}
Most of the literature on 2-dimensional conformal field theory (2d CFT) describes just what is called \emph{chiral conformal field theory}, formalized in terms of vertex operator
algebras or local conformal nets. 
But this only captures the holomorphic and low-genus aspect of conformal field
theory and is just one half of 
the data required for a full CFT, the remaining piece being the full solution of the
sewing constraints that 
makes the theory well defined for all genera.

With these questions -- fundamental as they are for perturbative string theory --
seemingly too hard to answer, 
a plethora of related model and toy model quantum field theoretic systems found
attention instead. A range of 
topological (quantum) field theories (T(Q)FTs) either approximates the physically
relevant CFTs as in the topological A-model and B-model, or encodes these holographically
in their boundary theory 
as for Chern-Simons theory and \emph{its} toy model, the Dijkgraaf-Witten theory. 

In this way a wealth of worldvolume QFTs appears that in 
some way or another is thought to encode information about string theory. 
Furthermore, in each case what
%
really matters is the \emph{full} worldvolume QFT: the rule that assigns correlators to all
possible worldvolume 
cobordisms, because this is what is needed 
%
even to write down the corresponding
second quantized 
perturbation series. However, despite this urgent necessity for understanding QFT on arbitrary cobordisms, 
the tools to study or even formulate this precisely were for a long time 
largely unavailable. 
Nevertheless, proposals for how to make these questions accessible to the
development of suitable 
mathematical machinery 
%
already existed.

Early on it was suggested, based on topological examples, that the path integral
and the state-propagation 
operators that it is supposed to yield are nothing but a representation of a category
of cobordisms \cite{Atiyah}. It was 
further noticed that this prescription is not restricted to TQFTs, and in fact
CFTs were 
proposed to be axiomatized as representations of categories of conformal cobordisms
\cite{Segal}.
In parallel to this development, another school developed a dual picture, now known as 
\emph{local}  or \emph{algebraic quantum field theory} (AQFT) \cite{Haag}, where 
it is not the state-propagation -- the
\emph{Schr{\"o}dinger picture} -- 
of QFT that is axiomatized and made accessible to high-powered machinery, 
but rather the assignment of algebras of 
observables -- the \emph{Heisenberg picture} of QFT. 

While these axiomatizations were known and thought of highly by a few select 
researchers who
worked on them, they 
were mostly happily ignored by the quantum field theory and string theory community at large, and to a good 
degree rightly so: nobody should trust an axiom system that has not yet 
proven its worth by providing 
useful theorems and describing nontrivial examples of interest. But neither the study of cobordism 
representations nor that of systems of algebras of observables could for a long time
-- apart from a few 
isolated exceptions -- claim to add much to the world-view of those who 
value formal structures in physics, 
but not a priori formal structures in mathematics. It is precisely this that is
changing now. 

Major structural results have been proven about the axioms of functorial quantum
field theory (FQFT)
in the form of cobordism representations and dually those of local nets of algebras
(AQFT) and factorization 
algebras. Furthermore,  classes of physically interesting examples have been constructed,
filling these axiom systems with life. 
We now provide a list of such results, which, while necessarily incomplete,
may serve to give an impression of the status of the field, and serve to put the contributions
of this book into perspective.


\medskip

\noindent {\bf I. Cobordism representations} 

\vspace{2mm}
\noindent{\bf (i) Topological case.} The most foundational result in TQFT is 
arguably the 
formulation and proof \cite{LurieTQFT} of the cobordism hypothesis 
\cite{BaezDolan} which classifies 
\emph{extended} (meaning: ``fully local'')  $n$-dimensional TQFT by the ``fully
dualizability"-structure on the 
``space'' of states (an object in a symmetric monoidal $(\infty,n)$-category) 
that it assigns to the point. 
(In \cite{SatiSchreiber} the contribution by Bergner \cite{ContributionBergner} surveys the formulation and proof of the
cobordism hypothesis).
This hugely facilitates the construction of
interesting   examples of extended $n$-dimensional TQFTs. For instance 
\begin{itemize}
\item recently it was understood  that 
the state-sum constructions of 3d TQFTs from fusion categories
(e.g. \cite{BakalovKirilov})
are subsumed by the cobordism hypothesis-theorem and the 
fact \cite{DSS} that fusion categories are among the fully dualizable objects in 
the $(\infty,3)$-category of monoidal categories with bimodule categories
as morphisms;

\item 
  the Calabi-Yau $A_\infty$-categories that Kontsevich conjectured 
  \cite{Kontsevich} encode the 2d TQFTs that participate
 in homological mirror symmetry have been understood to be the  ``almost fully dualizable'' objects (\emph{Calabi-Yau objects})
that classify extended open/closed 2-dimensional TQFTs on cobordisms with 
non-empty outgoing boundary with values in the $(\infty,1)$-category of chain complexes (``TCFTs'' \cite{Costello1}, \cite{LurieTQFT});

In this context crucial aspects of Witten's  observation in  \cite{WittenCS} have been made
precise \cite{Costello2}, relating Chern-Simons theory to the effective target space theory 
of the A- and B-model topological string, 
    thus providing a rigorous handle on an example of the effective background theory 
    induced by a string perturbation series 
    over all genera.
\end{itemize}

\medskip  
\noindent{\bf (ii)  Conformal case.} A complete classification of \emph{rational} full 2d CFTs
on cobordisms of all genera has been obtained in terms of Frobenius algebra objects in modular tensor
categories \cite{FRS}. While the
rational case is still 
``too simple" for the most interesting applications in string theory, 
its full solution shows that already 
here considerably more interesting structure is to be found than suggested by the naive
considerations in much 
of the physics literature.
(The contributions by Kapustin-Saulina \cite{ContributionKapustinSaulina} and by Kong \cite{ContributionKong} in \cite{SatiSchreiber} discuss aspects of this.)

\medskip
\noindent{\bf (iii) Supergeometric case.} There is now a full proof available, starting from 
the axioms, that the partition
function of a 
$(2|1)$-dimensional supersymmetric 2d-QFT indeed is a modular form, as suggested by
Witten's work \cite{WittenGenus} on the partition function of the heterotic string
and the index of the Dirac operator on loop space. (A formalization and proof of this fact 
in terms of supergeometric cobordism representations is described in the contribution by 
Stolz-Teichner \cite{ContributionStolzTeichner} to \cite{SatiSchreiber}.) This suggests a deep relationship between superstrings and 
the generalized cohomology theory called ${tmf}$ (for \emph{topological modular forms}) -- 
in a sense, the universal elliptic cohomology theory 
-- which lifts the more familiar relation between superparticles (spinors) and K-theory
to higher categorical dimension. (This is the content of the contribution by Douglas-Henriques \cite{ContributionDouglasHenriques}
in \cite{SatiSchreiber}.)

\medskip
\noindent{\bf (iv) Boundary conditions and defects/domain walls.}
One simple kind of extra structure on cobordisms that is of profound importance 
is boundary labels and decompositions of cobordisms into domains, meeting at \emph{domain walls} (``defects''). 
(The definition of QFT with defects is part of the content of the contribution by
Davydov-Runkel-Kong \cite{ContributionDavydovKongRunkel} to \cite{SatiSchreiber}).
That cobordism representations with boundaries for the string encode 
D-branes on target space was originally amplified by Moore and Segal \cite{MooreSegal}. Typically open-closed QFTs are entirely determined by 
their open sectors and boundary conditions,
a fact that via \cite{Costello1}
led to Lurie's proof of the cobordism hypothesis. 
(A survey of a list of results on presentation of 2d CFT by algebras of boundary data
is in the contribution by Kong \cite{ContributionKong} to \cite{SatiSchreiber}.)

\medskip
\noindent{\bf  (v) Holographic principle.} 
A striking aspect of the classification of rational 
CFT mentioned above is that it proceeds -- rigorously -- by a version of the 
\emph{holographic principle}. This states that under some conditions the
partition function and correlators of an $n$-dimensional QFT are encoded in the
\emph{states} of an $(n+1)$-dimensional TQFT in codimension 1. The first
example of this had been the holographic relation between 3-dimensional 
Chern-Simons theory and the 2-dimensional WZW CFT in the seminal work
\cite{WittenJonesPolynomial}, which marked the beginning of the investigation of
TQFT in the first place. A grand example of the principle is the 
AdS/CFT conjecture, which states that type II string theory itself is holographically
related to super Yang-Mills theory. While mathematical formalizations of AdS/CFT
are not available to date, lower dimensional examples are finding precise formulations.
(The contribution by Kapustin-Saulina \cite{ContributionKapustinSaulina} in \cite{SatiSchreiber} discusses how the 
construction of rational 2d CFT by \cite{FRS} is naturally induced from 
applying the holographic principle to Chern-Simons theory with defects).

One of the editors once suggested that, in the formalization by 
cobordism representations, holography corresponds to the fact that 
\emph{transformations} between $(n+1)$-functors are in components themselves
essentially given by $n$-functors. A formalization of this observation
for extended 2d QFT has been given in \cite{SchommerPriesDefects}.
(The contribution by Stolz-Teichner \cite{ContributionStolzTeichner} to \cite{SatiSchreiber} 
crucially uses transformations between higher dimensional
QFTs to \emph{twist} lower dimensional QFTs.)

\newpage
\noindent {\bf II. Systems of algebras of observables}

\vspace{2mm}
\noindent{\bf (i) Nets of algebras.} In the form of the \emph{Haag-Kastler axioms}, the description of QFT through its local algebras of 
observables had been given a clean mathematical formulation
\cite{HalvorsonMueger} a long time ago \cite{HaagKastler}.
This approach had long produced fundamental 
structural results about QFT, 
such as the PCT theorem and the spin-statistics theorem
(cf. \cite{SW}). 
Only recently has it finally been shown in 
detail \cite{BrunettiDuetschFredenhagen}
how examples of AQFT nets can indeed be constructed along the lines
of perturbation theory and Wilsonian effective field theory, thus connecting 
the major tools of practicing particle physicists with one of the major 
formal axiom systems.
Using an operadic variant of Haag-Kastler nets in the case of Euclidean 
(``Wick rotated'') QFT -- called \emph{factorization algebras} -- 
a similar discussion is sketched in \cite{CostelloGwilliam}. 
At the same time, the original axioms have been found to naturally generalize
from Minkowski spacetime to general (globally hyperbolic) curved and topologically nontrivial spacetimes \cite{BFV}.

\medskip
\noindent{\bf (ii) Boundaries and defects.}
The Haag-Kastler axioms
had been most fruitful in the description of 2 dimensional and conformal
field theory (``conformal nets''), where they serve to classify chiral 2d CFTs
\cite{KawahigashiLongo}\cite{Kawahigashi}, construct integral 2d QFTs \cite{Lechner} 
and obtain insights into boundary field theories (open strings) \cite{LongoRehren}. 
Remarkably, the latter has recently allowed a rigorous re-examination
\cite{LongoWitten} of old arguments about the background-independence of 
string field theory. 
(The contribution by Douglas-Henriques in \cite{SatiSchreiber} presents a modern
version of the Haag-Kastler axioms for conformal nets and extends the discussion
from boundary field theory to field theory with defects.)

\medskip
\noindent{\bf (iii) Higher chiral algebras.} 
The geometric reformulation of vertex operator algebras 
in terms of \emph{chiral algebras} \cite{BD} has proven to be fruitful, 
in particular in its higher categorical generalizations 
\cite{LurieAlgebra} by factorizable cosheaves of $\infty$-algebras.
While the classical AQFT school restricted attention to QFT over trivial
topologies, it turns out that also topological QFTs can be described and constructed by 
local assignments of algebras ``of observables''. In \cite{LurieTQFT}   
$n$-dimensional extended  TQFTs are constructed
from \emph{$E_n$-algebras} -- algebras over the little $n$-cubes operad 
-- by a construction called
\emph{topological chiral homology}, which is a grand generalization of 
Hochschild homology over arbitrary topologies.
(The contribution by Weiss \cite{ContributionWeiss} in \cite{SatiSchreiber} discusses the theory of homotopy algebras
over operads involved in these constructions.)

This last work is currently perhaps the most formalized and direct bridge between the two 
axiom systems, the functorial and the algebraic one.
This indicates the closure of a grand circle of ideas and makes the outline
 of a comprehensive fundamental formalization of full higher-genus QFT
  visible. 

\medskip

\noindent {\bf III. Quantization of classical field theories} 

\vspace{2mm}
While a realistic axiomatization is the basis for all mathematical progress in 
QFT, perhaps even more important in the long run for physics is that with the supposed 
\emph{outcome} of the (path integral) quantization process thus identified precisely
by axioms for QFT, 
it becomes possible to consider the nature of the \emph{quantization} process itself. This is  particularly relevant in applications of QFT as worldvolume theories in string
theory, 
where one wishes to explicitly consider QFTs 
that arise as the quantization of
sigma-models with specified gauge background fields. A good understanding of this quantization
step is one of the links between the worldvolume theory and the target space theory 
and hence between the abstract algebraic description of the worldvolume QFT and the phenomenological interpretation of its correlators  in its target space, ultimately connecting theory to experiment.
We now indicate some of the progress in mathematically understanding the process of 
quantization in general and of sigma-models in particular.

\medskip
\noindent{\bf (i) Path integral quantization.}
It has been suggested (e.g. \cite{FreedPathIntegral}) that the path integral is
to be understood abstractly as a pull-push operation -- an \emph{integral transform} --
acting on states in the form of certain cocycles, by first pulling them up to the space of worldvolume configurations along the map induced by the incoming boundary, and then pushing forward along the map induced by the outgoing boundary. This is fairly well understood for Dijkgraaf-Witten theory \cite{FreedQuinn}.  In \cite{FreedHopkinsLurieTeleman} it is claimed that at least for
all the higher analogs of Dijkgraaf-Witten theory (such as the Yetter model 
\cite{MartinsPorter})
a formal pull-push path integral quantization procedure exists in terms of colimits of 
$n$-categorical algebras, yielding fully extended TQFTs.

A more geometric example for which pull-push quantization is well understood
is Gromov-Witten theory \cite{Katz}. More recently also Chas-Sullivan's string topology operations have been understood this way, for strings on a single brane in \cite{Godin} and recently for arbitrary branes in \cite{Kupers}. In \cite{BenZviNadler} it is shown that such integral transforms exist on stable $\infty$-categories of quasicoherent sheaves for all target spaces that are perfect derived algebraic stacks, each of them thus yielding a 2-dimensional TQFT from background geometry data. 


\medskip
\noindent{\bf (ii) Higher background gauge fields.}
Before even entering (path integral) quantization, there is a fair bit of mathematical
subtleties involved in  the very definition of the string's action functional in the
term that describes the coupling to the higher background gauge fields,
such as the Neveu-Schwarz (NS) $B$-field and the Ramond-Ramond (RR) fields. 
All of these are recently being understood
systematically in terms of \emph{generalized differential cohomology} \cite{HopkinsSinger}.	

Early on it had been observed that the string's coupling to the  B-field
is globally occurring via the higher dimensional analog of the line
holonomy of a circle bundle: the surface holonomy 
\cite{GawedzkiReis}\cite{FNSW} of a \emph{circle 2-bundle with connection}  
\cite{Schreiber}: a \emph{bundle gerbe} with connection, classified
by degree-3 ordinary differential cohomology. 
More generally, on orientifold target space 
backgrounds it is the nonabelian $(\mathbb{Z}_2//U(1))$-surface holomomy
\cite{SchreiberWaldorf}\cite{Nikolaus}
over unoriented surfaces \cite{SSW}.

After the idea had materialized that the RR fields have to be 
regarded in K-theory
\cite{MW} \cite{FreedHopkins},
it eventually became clear \cite{FreedDiracQuantization} that all
the higher abelian background fields appearing in the effective supergravity theories of string theory are properly 
to be regarded as cocycles in 
\emph{generalized differential cohomology} \cite{HopkinsSinger} 
-- the RR-field being described by 
  \emph{differential K-theory} \cite{BunkeSchick} --
and even more generally in \emph{twisted} such theories:
the presence of the B-field makes the RR-fields live in \emph{twisted K-theory}
(cf. \cite{BMRZ}).

A perfectly clear picture of twisted generalized cohomology theory in terms of associated
$E_\infty$-module spectrum $\infty$-bundle has been given in 
\cite{AndoBlumbergGepner}. This article in particular identifies the twists of ${tmf}$-theory, 
which are expected \cite{Sati} \cite{AndoSati} to play a role in M-theory in 
the higher analogy of twisted K-theory in string theory.

\medskip
\noindent{\bf (iii) Quantum anomaly cancellation.}
The cancellation of the quantum anomaly of fermions on the superstring's worldvolume -- the (differential) class of their Pfaffian line bundles on the bosonic configuration space -- imposes subtle conditions on the background gauge fields on spacetime to which the string couples.

By means of the machinery of generalized differential cohomology, 
recently \cite{Bunke} makes fully precise the old argument of Killingback about the worldsheet version of the
celebrated \emph{Green-Schwarz anomaly cancellation} (the effect that initiated the ``First superstring revolution''), using a model for \emph{twisted differential string structures} 
\cite{SSSIII} \cite{FSS}  in terms of bundle gerbes, due to \cite{Waldorf}.
These differential string structures -- controlled by the higher Lie and Chern-Weil theory of the \emph{smooth string 2-group} \cite{Henriques}\cite{BCSS}
-- are the higher superstring analogs in higher smooth geometry \cite{Schreiber} of the spin-bundles with connection that control the dynamics of spinning/superparticles. 

(In \cite{SatiSchreiber} the contribution by Distler-Freed-Moore \cite{ContributionDistlerFreedMoore} presents what is to date the most accurate description of the conditions on the differential cohomology classes of the superstring's background gauge fields for general orbifold and orientifold target spaces.)

\vspace{.5cm}
Taken together, all these developments should go a long way towards understanding the
fundamental nature of QFT on arbitrary cobordisms and of the string perturbation 
series defined by such 2d QFTs. 
However, even in the light of all these developments, the
reader accustomed to the 
prevailing physics literature may still complain that none of this progress in QFT
on cobordisms of all 
genera yields a definition of what string theory really is. Of course this is true
if by ``string theory" 
one understands its non-perturbative definition. But this supposed non-perturbative
definition of string theory is beyond reach at the moment. 
Marvelling -- with a certain admiration of their audacity -- at how ill-defined
this is has made 
the community forget that something much more mundane, 
the perturbation series over  CFT correlators that defines \emph{perturbative string
theory}, 
has been ill-defined all along: only the machinery of full CFT in terms of cobordism
representations gives a precise 
meaning to what exactly it is that the string pertubation series is a series over.
Perhaps it causes feelings of 
disappointment to be thrown back from the realm of speculations about
non-perturbative string theory to just the 
perturbation series. But at least this time one lands on solid ground, which is the
only ground that serves as 
a good jumping-off point for further speculation.

In string theory it has been the tradition to speak of major conceptual insights
into the theory as 
\emph{revolutions} of the theory. The community speaks of a first and a second
superstring revolution 
and a certain longing for the third one to arrive can be sensed. With a large part
of the community busy 
attacking grand structures with arguably insufficient tools, it does not seem 
farfetched that 
when the third one does arrive, it will have come out of 
mathematics departments. \footnote{See in this context for instance the 
opening and closing talks at 
the \href{http://media.medfarm.uu.se/flvplayer/strings2011/}{\emph{Strings 2011}} conference.}

\medskip
\medskip

\newpage

\begin{center}
{\Large \bf Selected expositions}
\end{center}

In \cite{SatiSchreiber} we have collected selected expositions by researchers in the field that discuss aspects of the kind of developments that we have described here. Here we outline the content of that volume, highlighting 
how the various articles are related and emphasizing
how they fit into the big picture that we have drawn above.

\medskip
\noindent {\bf I. Foundations of Quantum Field Theory}

\medskip
\noindent{\bf 1. Models for $(\infty,n)$-Categories and the Cobordism Hypothesis} 
-- by {\it Julia Bergner} \cite{ContributionBergner}.

The Schr{\"o}dinger picture of extended topological quantum field theory of dimension $n$ 
is formalized 
as being an $(\infty,n)$-functor on the $(\infty,n)$-category of cobordisms of
dimension $n$.
This article reviews the definition and construction of the ingredients of this
statement, due to \cite{LurieTQFT}.

This picture is the basis for the formulation of QFTs on cobordisms with structure. Contributions below discuss cobordisms with defect structure, with conformal structure and with flat Riemannian structure.

\medskip
\noindent{\bf 2. From operads to dendroidal sets} -- by {\it Ittay Weiss} \cite{ContributionWeiss}.

The higher algebra that appears in the algebraic description of QFT 
-- by local nets of observables,
factorization algebra or chiral algebras --
 is in general operadic. For instance the vertex operator algebras appearing in the
description of CFT 
(see Liang Kong's contribution below) are algebras   over an operad of holomorphic
punctured spheres.

This article reviews the theory of operads and then discusses a powerful
presentation in terms of 
dendroidal sets -- the operadic analog of what simplicial sets are for
$(\infty,1)$-categories.
This provides the homotopy theory for $(\infty,1)$-operads, closely related to the 
traditional model by topological operads.

\medskip
\noindent{\bf 3. Field theories with defects and the centre functor}  -- by 
{\it Alexei Davydov, Liang Kong and Ingo Runkel} \cite{ContributionDavydovKongRunkel}. 

This article gives a detailed discussion of cobordism categories for cobordisms 
with defects/domain walls. 
An explicit construction of a lattice model of two-dimensional TQFT with defects is
spelled out. The authors isolate 
a crucial aspect of the algebraic structure induced by defect TQFTs on their spaces
of states: as opposed to the 
algebra of ordinary bulk states, that of defect states is in general
non-commutative, but certain worldsheet 
topologies serve to naturally produce the centre of these algebras.

Below in \emph{Surface operators in 3d TQFT} topological field theories with defects
are shown to induce, by 
a holographic principle, algebraic models for 2-dimensional CFT. 
In \emph{Topological modular forms and conformal nets} conformal 
field theories with defects are considered.

\medskip
\medskip
\noindent  {\bf II. Quantization of Field Theories}

\medskip
\noindent{\bf 1. Homotopical Poisson reduction of gauge theories} -- by 
{\it Fr{\'e}d{\'e}ric Paugam} \cite{ContributionPaugam}.

The basic idea of quantization of a Lagrangian field theory is simple: one forms the
covariant phase space given 
as the critical locus of the action functional, then forms the quotient by gauge
transformations 
and constructs the canonical symplectic form.
Finally, one applies 
deformation quantization or geometric quantization to the resulting symplectic manifold.

However, to make this naive picture work, 
care has to be taken to form both the intersection
(critical locus) and the 
quotient (by symmetries) not naively but up to homotopy in 
derived geometry \cite{LurieStructured}. The
resulting derived covariant 
phase space is known in physics in terms of its
Batalin-Vilkovisky--Becchi-Rouet-Stora-Tyutin (BV-BRST) 
complex. This article reviews the powerful description of variational calculus and
the construction of the 
BV-BRST complex in terms of D-geometry \cite{BD} -- the geometry over de Rham spaces -- and
uses this to 
analyze subtle finiteness conditions on the BV-construction.

\medskip
\noindent{\bf 2. Orientifold pr{\'e}cis}  -- by 
{\it Jacques Distler, Daniel Freed, and Gregory Moore} \cite{ContributionDistlerFreedMoore}.

The consistent quantization of the sigma model for the (super-)string famously
requires the target space 
geometry to satisfy the 
Euler-Lagrange equations of an effective supergravity theory on target space.
In addition there are subtle cohomological conditions for the cancellation of fermionic worldsheet
anomalies.

This article discusses the 
intricate conditions on the differential cohomology of the background fields --
namely the Neveu-Schwarz  $B$-field in ordinary differential cohomology 
(or a slight variant, which the authors discuss) and the RR-field 
in differential K-theory twisted by the $B$-field -- in particular if target space
is allowed to be not 
just a smooth manifold but more generally an orbifold and even more generally an
orientifold. Among 
other things, the result shows that the ``landscape of string theory vacua'' --
roughly the moduli space 
of consistent perturbative string backgrounds (cf. \cite{Do}) -- is more 
subtle an object than often assumed in the literature.

\medskip
\medskip
\noindent {\bf III. Quantum two-dimensional Field Theories}

\medskip
\noindent{\bf 1. Surface operators in 3d TFT and 2d Rational CFT}  -- by 
{\it Anton Kapustin and Natalia Saulina} \cite{ContributionKapustinSaulina}. 

Ever since Witten's work on 3-dimensional Chern-Simons theory it was known
that by a holographic principle this theory 
induces a 2d CFT on 2-dimensional boundary surfaces. 
This article amplifies that if one thinks of the 3d Chern-Simons TQFT as a
topological QFT with defects, then the structures formed by codimension-0 defects bounded by
codimension-1 defects naturally reproduce, 
holographically, the description of 2d CFT by Frobenius algebra 
objects in modular tensor categories \cite{FRS}.

\medskip
\noindent{\bf 2. Conformal field theory and a new geometry}  -- by {\it Liang Kong} \cite{ContributionKong}.

While the previous article has shown that the concept of TQFT together with the
holographic principle naturally imply that 
2-dimensional CFT is encoded by monoid objects in modular tensor categories, this
article reviews a series of strong 
results about the details of this encoding. 
In view of these results and since every 2d CFT also induces an effective target
space geometry -- as described in more 
detail in the following contribution -- the author amplifies the fact that stringy
geometry is thus presented 
by a categorified version of the familiar duality
between spaces and algebras: now for 
algebra objects internal to suitable monoidal categories.

\newpage
\noindent{\bf 3. Collapsing Conformal Field Theories, 
spaces with non-negative Ricci curvature and  non-commutative geometry}  -- by {\it Yan Soibelman} \cite{ContributionSoibelman}.

The premise of perturbative string theory is that every 
suitable 2d (super-)CFT describes the quantum sigma model for a string propagating
in \emph{some} target 
 space geometry, if only we understand this statement in a sufficiently general context of
geometry, such as spectral noncommutative geometry.
In this article the author analyzes the geometries induces from quantum strings in the 
point-particle limit (``collapse limit") where only the lowest string excitations are
relevant. In the limit the 
algebraic data of the SCFT produces a spectral triple, which had been shown by Alain
Connes to encode generalized 
Riemannian geometry in terms of the spectrum of Hamiltonian operators. The author
uses this to demonstrate compactness results 
about the resulting moduli space of ``quantum Riemann spaces''.

\medskip
\noindent{\bf 4. Supersymmetric field theories and generalized cohomology} -- by {\it Stephan Stolz and Peter Teichner} \cite{ContributionStolzTeichner}.

Ever since Witten's derivation of what is now called the \emph{Witten genus} as the
partition function of the heterotic superstring, there have been indications that
superstring physics should be governed by the generalized cohomology theory called
topological modular forms ($tmf$)  in analogy to how super/spinning point particles
are related to K-theory. 
In this article the authors discuss the latest status of their seminal program of
understanding these cohomological phenomena
from a systematic description of functorial 2d QFT with metric structure on the cobordisms.

After noticing that key cohomological properties of the
superstring depend only on supersymmetry and not actually on conformal invariance, 
the authors
simplify to cobordisms with flat super-Riemannian structure, but equipped with maps
into some auxiliary target space $X$. A classification of such
QFTs by generalized cohomology theories on $X$ is described: 
a relation between $(1|1)$-dimensional flat Riemannian field theories and 
K-theory and between $(2|1)$-dimensional flat Riemannian 
field theories and $tmf$.

\medskip
\noindent{\bf 5. Topological modular forms and conformal nets}  -- by 
{\it Christopher Douglas and Andr{\'e} Henriques} \cite{ContributionDouglasHenriques}.

Following in spirit the previous contribution, 
but working with the AQFT-description
instead, the authors of this article describe a refinement of 
conformal nets, hence of 2d CFT, 
incorporating defects. Using this they obtain a tricategory of fermionic conformal nets
(``spinning strings'') which constitutes a higher analog of the bicategory of Clifford algebras.
Evidence is provided which shows that these \emph{categorified} spinors are related to
$tmf$ in close analogy to how ordinary Clifford algebra is related to K-theory,
providing a concrete incarnation of the principle by which string physics is a form of 
categorified particle physics.

\vspace{1.2cm}
\noindent {\bf Acknowledgements.}
 The authors would like to thank Arthur 
Greenspoon for his very useful editorial input on this introduction as well as on the 
papers in the volume.


\renewcommand\refname{Contributions to \cite{SatiSchreiber}}

\end{document}